\documentclass[aps,eqsecnum,amsmath,twocolumn,preprintnumbers]{revtex4-1}
\usepackage{graphics,graphicx,setspace,epsfig,color}
\usepackage[letterpaper, dvips,width=7.5in,height=8.5in,includemp=false]{geometry}
\usepackage[vcentermath]{}
\usepackage{epsfig}
\usepackage{amscd}
\usepackage{amssymb,amsmath}
\usepackage{url}

\newcommand{\nn}{\nonumber}
\newcommand{\figref}[1]{Fig.~\ref{#1}}
\newcommand{\tableref}[1]{Table~\ref{#1}}

\newcommand{\ket}[1]{|#1\rangle}

\begin{document}

\title{Stable vortex loops in two-species BECs}

\author{Paulo F. Bedaque}
\author{Evan Berkowitz}
\author{Srimoyee Sen}
\affiliation{University of Maryland\\
Department of Physics\\
College Park,  MD 20742-4111}

\preprint{UM-DOE/ER/40762-510}

\begin{abstract}
We consider the creation of stable, stationary closed vortex loops, analogue to the vortons and superconducting cosmic strings, in cold atom BEC's. We explore the parameter region where these solutions are likely to exist and comment on methods to create them experimentally.
\end{abstract}

\maketitle
\section{Introduction}
The existence of vortex solutions is one of the hallmarks of superfluidity. They are seen in superfluid ${}^4$He as well as in Bose-Einstein condensates in cold atom traps \cite{Abo-Shaeer20042001}. Superfluid vortices are also the main actors in the leading theory of rotation and ``glitches'' observed in rotating neutron stars (pulsars)\cite{ISI:A1969E722500025}; superfluid vortices are supported in both the hadron and quark matter phases at high density. Similar structures are also likely to  appear in Grand Unified Theories (GUT) of physics beyond the standard model of particle physics. In that context vortex filaments in the early universe might serve as seeds around which galaxies could form\cite{Hindmarsh:1994re}.

The stability of superfluid vortices is guaranteed by their topological properties. The velocity circulation around a vortex is characterized by a phase which wraps axially around the vortex.  Because the field must be single-valued this phase must go through an integer number of cycles and thus cannot change continuously. Discontinuous changes through quantum tunneling or thermal activation are usually negligible.  

Vortices with opposite circulation can annihilate one another. This phenomenon seems to preclude the possibility of stable closed vortex loops, because the opposite sides of a closed loop form a vortex/antivortex pair that can annihilate and, in fact, the energetics of the system favors annihilation. Because a vortex has a tension $T$ along its length, the energy of a vortex loop is proportional to its length: $E\sim 2\pi R T$, where $R$ is the radius of the loop.  This energy is reduced as the loop shrinks. When the opposite sides of the loop come closer than their thickness they can annihilate. Single-species vortex loops can only be stabilized by the Magnus force, which will sustain a loop if it is moving at a specific velocity in a direction perpendicular to its plane.
\begin{figure}[t]
\includegraphics[width=\columnwidth]{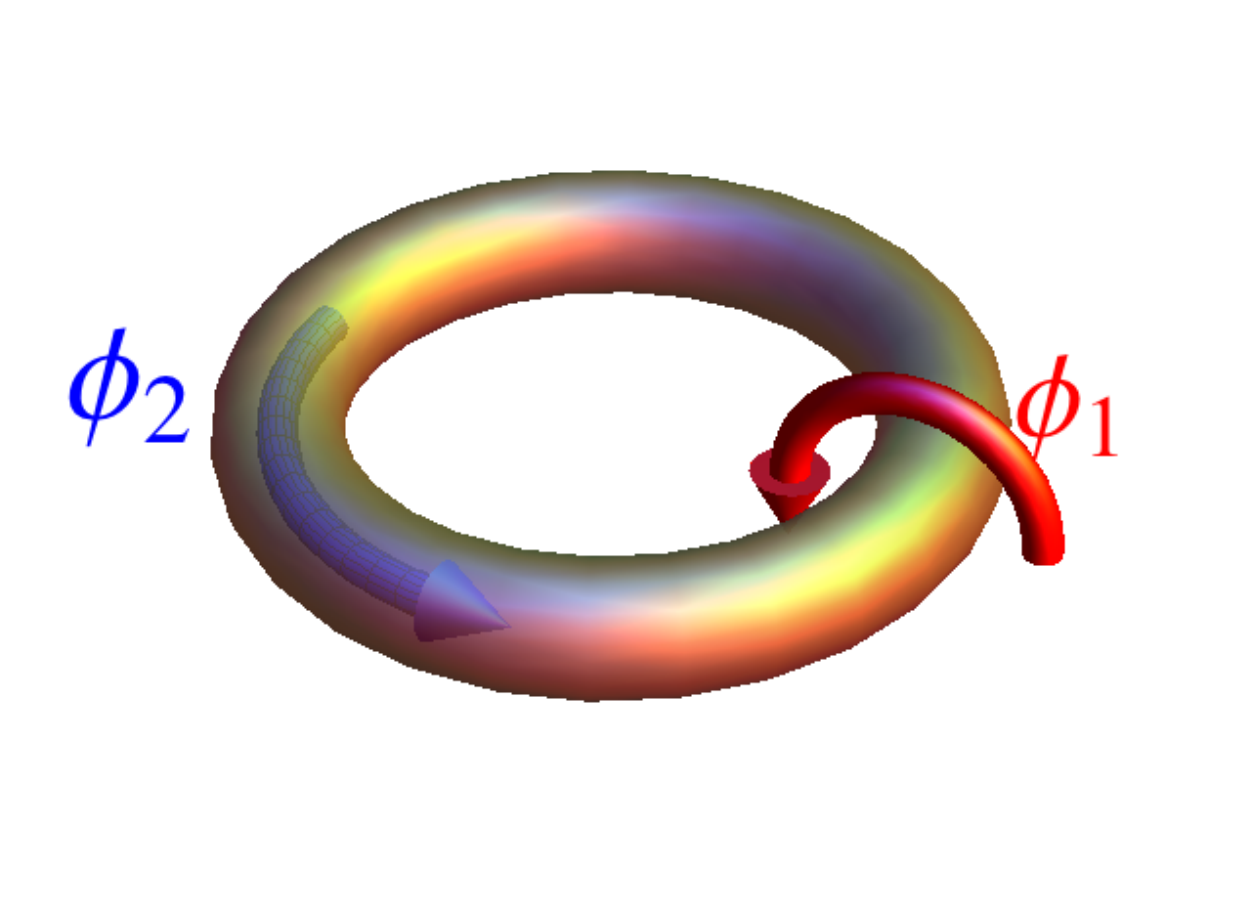}
\caption{The arrows denote the momentum (change in phase) of species 1 and 2.}
\label{fig:vorton}
\end{figure}

It has been pointed out in the context of relativistic models that a stable closed vortex loop can exist if two species are competing to condense \cite{Davis1989209}. Qualitatively, the mechanism is straightforward. Suppose in some region there is a bulk where the first species was condensed, and there is a vortex in that medium.  In that vortex's core the condensate of species 1 vanishes. If the two species repel each other it will be energetically favorable for the second species to condense in the core of the vortices of species 1 rather than occupy the same space as the first species.  Suppose that the vortex of species 1 forms a closed loop and, in  addition, there is a non-zero vorticity of species 2 along the core of the loop (see \figref{fig:vorton}). The energy cost related to the vorticity $l$ of species 2 is proportional to $2\pi R (l/R)^2 \sim l^2/R$ expresses the kinetic energy of species 2 inside the vortex. Because this contribution increases as the loop shrinks, there will always be some nonzero radius where the total energy $E\sim T R + l^{2}/R$ has a minimum.  Thus, if this equilibrium radius is big enough to preclude annihilation, the second species provides a stabilization mechanism for vortex loops.

If the particle condensing in the {\it interior} of the vortex is charged, the vortex will be superconducting and solutions of this kind are known as superconducting strings \cite{Witten:1984eb} in order to distinguish them from superconducting flux tubes (Abrikosov vortices) where the superconducting region is {\it outside} of the vortex. Closed loops of superfluid vortices stabilized by the mechanism sketched above are know as ``vortons''. Vortons can arise in GUT-scale models during the early universe\cite{Davis:1988ij}. Quark matter, which can exist in the core of neutron stars, also provides the ingredients for the existence of vortons, the role of the two competing species played by quasiparticle excitations of the color superconducting ground state with quantum numbers of neutral and charged kaons \cite{Kaplan:2001hh,Buckley:2002mx,Buckley:2002ur,Bedaque:2011fg}.

The purpose of this paper is to study the possibility that stable vortons can exist in cold atomic traps. The (meta)stability of this kind of solution is analyzed in the first part through a calculation of the energy of the vorton as a function of its radius and thickness. We first show that in some circumstances, the equilibrium radius of the vorton is on the scale of microns and is reasonably larger than its thickness. We then comment on  possible ways of actually creating vortons experimentally.

\section{Stability and equilibrium properties of vortons}
Consider a system of two species of spinless bosons. In the dilute limit, where  the interparticle distances are typically larger than the scattering lengths, this system can be described by the Hamiltonian
\begin{equation}\label{eq:hamiltonian}
H=\frac{\hbar^2}{2M_1}|\nabla\phi_1|^2+\frac{\hbar^2}{2M_2}|\nabla\phi_2|^2+V(\phi_1,\phi_2),
\end{equation}
where
\begin{equation}\label{int}
\begin{split}
V(\phi_1,\phi_2)=\frac{1}{2}\frac{8\pi\hbar^2 a_1}{M_1}|\phi_1|^4+\frac{1}{2}\frac{8\pi\hbar^2 a_2}{M_2}|\phi_2|^4\\
+\frac{2\pi\hbar^2 a_{12}}{M_{12}}|\phi_1|^2|\phi_2|^2-\mu_1 |\phi_1|^2.
\end{split}
\end{equation}
Here, $\phi_1$ and $\phi_2$ are the second quantized fields annihilating species 1 and 2, $a_{i}$ and $M_{i}$ are the scattering lengths and masses of the two species ($i$=1,2), $a_{12}$ is the interspecies scattering length, and $M_{12}$ is the reduced mass of the two species. The chemical potential of species 1 is given, in the absence of species 2 and at leading order in the diluteness expansion, by
\begin{equation}\label{eq:mu1}
\mu_1=\frac{8\pi\hbar^2n_1 a_1}{M_1}
\end{equation}
where, $n_1$ is the asymptotic density of the first species.  We will always be interested in situations where species 1 forms the bulk material, so we now eliminate $\mu_{1}$ from the discussion.  We do not include a chemical potential term for species 2; instead we will work with a fixed number of particles of that type. We will be interested in the phase separation  regime \mbox{$4 a_1 a_2/M_1 M_2 < \left(a_{12}/M_{12}\right)^2$,} where the interspecies repulsion encourages the two species to stay at separate points in space \cite{pethick2002bose}.  

Bose-Einstein condensation is described by a non-vanishing matrix element the field operators $\phi_{1,2}$. At low densities ($n_{1,2}^{-1/3}\gg a_1, a_2, a_{12}$) the mean field approximation is valid and the matrix element (also denoted by $\phi_{1,2}$) satisfy the classical equations of motion, that is, the Gross-Pitaevski equations. For a  straight vortex of only the first species, $\phi_1$ is of the form
\begin{equation}
\phi_{1}(r,\theta,z) = f_1(r) e^{i j \theta},
\end{equation} 
where $r$, $\theta$, and $z$ are cylindrical coordinates, $f_{1}(r=0)=0$ and $f_{1}(r\rightarrow\infty) = \sqrt{n_1}$.  To keep $\phi_{1}$ single valued the winding number $j$ must be an integer.  We always use $j=1$, as vortices with larger $j$ are unstable and split into $j$ vortices with winding number 1, but keeping general $j$ is useful for studying the energy of a vorton. The precise profile $f_1(r)$ can be obtained by solving the Gross-Pitaevski equations or, equivalently, minimizing the energy. 

Simple scaling arguments imply that the solution $f_1(r)$ will change from $f_1=0$ to $f_1=\sqrt{n_1}$ over a distance of order $\delta\approx 1/\sqrt{8\pi n_1 a_1}$, the string thickness. The toroidal geometry of the vorton makes an analogous calculation a little involved. We can, however, bypass most of the difficulty by assuming that the radius $R$ of the vorton is much larger than its thickness $\delta$. In this case we can compute the energy of the vorton by computing the energy per length (that is, the tension) of a straight vortex and multiplying it by the length $2\pi R$. This approximation neglects the energy associated with the curvature of the vorton. For a straight vortex of the first species with an internal current of the second, the mean-field solution has the form
\begin{align}\label{eq:phi1phi2}
\phi_1&=f_{1}(r)e^{i j \theta},\nn\\
\phi_2&=f_{2}(r) e^{ikz}.
\end{align}

To encode the fact that we wish to consider a vortex loop of radius $R$, we impose periodic boundary conditions in $z$, identifying $z=\pm \pi R$, so that $z$ spans the arclength of the loop.  The continuity of the phase of $\phi_2$ implies that $k=l/R$, with integer $l$.

Plugging the functional forms in \eqref{eq:phi1phi2} with $k=l/R$ into the Hamiltonian given in \eqref{eq:hamiltonian}, the total energy of a vorton of a radius $R$ is
\begin{align}\label{eq:energy}
E\approx& 2\pi R \int dr \  2\pi r \left[\frac{\hbar^2}{2M_1}\left(\left(\frac{\partial f_1}{\partial r}\right)^2+\frac{ j^{2}}{r^2}f_{1}(r)^{2}\right)\right.\\
& \left.
+\frac{\hbar^2}{2M_2}\left(\left(\frac{\partial f_2}{\partial r}\right)^2
+\frac{l^2}{R^2} f_2 ^2\right)+V\left(f_1,f_2\right)+\frac{1}{2}\frac{8\pi\hbar^2 a_1n_1^2}{M_1}\right]\nonumber
\end{align}
which becomes more accurate as the ratio $R/\delta$ grows.  One might worry that including curvature corrections might encourage the vorton's radius to be smaller.  Brief consideration alleviates this fear: the configuration we consider when formulating \eqref{eq:energy} has no gradients of $\phi_{1}$ along the vorton's length.  Curvature effects will make those terms nonzero, and the nonuniformity of $\phi_{1}$ will provide a considerable potential barrier.  Therefore we expect that calculation that included curvature effects would find an $R$ larger than the $R$ we find from considering the energy expression in \eqref{eq:energy}.

In \eqref{eq:energy} we measure the energy in relation to the homogeneous ground state $j=0$, $f_1=\sqrt{n_1}$, $f_{2}=0$.  These parameters correspond to no vorton at all.  Any vorton will have a greater energy, and thus must be at best metastable.

We now argue that vortons are indeed long-lived.  If a vorton is initially created with a non-zero number $N_2$ particles of species 2 in its interior, we hope they remain trapped there, or the stabilizing energy that scales inversely with $R$ will disappear, causing the vorton's collapse.  Escaping the vorton would require these particles to go through a large region where species 1 is condensed, and that is energetically expensive.  One could also imagine a bubble of species 2 detaching from the vorton and moving towards the edge of the bulk, but that costs an amount of energy proportional to the area of the bubble due to its surface tension.  These escape mechanisms are costly, and thus we treat $N_{2}$ as a conserved quantity that is confined to the vorton's interior.  To implement this conservation, when we minimize the energy, we keep
\begin{equation}\label{eq:N2}
N_2 = 2\pi R\int dr\ 2\pi r |f_2(r)|^2
\end{equation} fixed. 

\begin{figure}[t]
\includegraphics[width=\columnwidth]{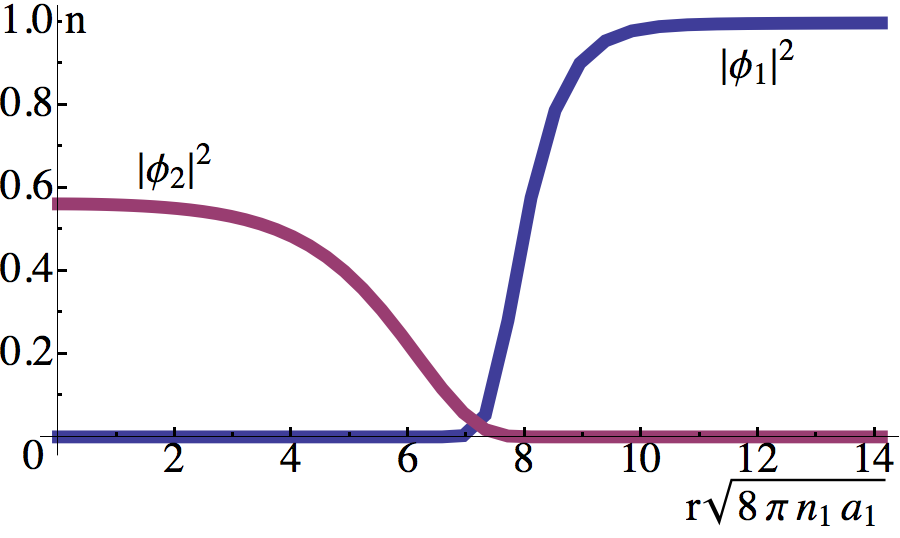}
\caption{The radial density profile of the two species $^{7}${Li}  and $^{87}${Rb} and $l=5$ The remaining parameters are given in the first row of \tableref{tab:data}.}
\label{fig:LiRb}
\end{figure}

To summarize, we find our vorton candidates by numerically minimizing  the energy in \eqref{eq:energy} in relation to $f_1(r), f_2(r)$ and $R$ while keeping the parameters $N_{2}$, $n_1$, $a_1$, $a_2$, $a_{12}$, $M_1$ and $M_2$ fixed.  We have eliminated $\mu_{1}$, and henceforth take $j=1$.  A typical example of the profile functions $f_1(r)$ and $f_2(r)$ resulting from this minimization is shown in \figref{fig:LiRb}.

It is easy to foresee some qualitative dependences before doing the numerical minimization. For instance, the larger the species 2 vorticity $l$ is, the larger $R$ and $R/\delta$ should be. A larger value of $a_{12}$ increases interspecies repulsion and shrinks the region where both species coexist. This tendency should quickly saturate, as the coexistence region becomes negligible and $a_{12}$ becomes effectively infinite.  We can also see that if we increase $n_1$ by a factor of $\eta$ and reduce the number of particles of species 2 by a factor of $\sqrt{\eta}$, both the radius and thickness decrease by a factor of $\sqrt{\eta}$ and therefore $R/\delta$ remains unchanged. 

On the other hand, the dependence of $R/\delta$ on some parameters are harder to anticipate.  For example, as  $M_1/M_2$ grows, the terms in the energy describing species 2 (and its interaction with species 1) grow. Because species 2 exerts a pressure counterbalanced by the bulk, this increase of the contribution from species 2 should increase both $R$ and $\delta$, so it is difficult to know in advance if their ratio will grow or shrink. Numerical studies indicate that the ratio $R/\delta$ increases with $M_{1}/M_{2}$. We also observe an increase of $R/\delta$ with increasing $a_1$ and decreasing $a_2$.

To make our detailed discussion more concrete, we study the vortons formed when the roles of species 1 and 2 are played by the $\ket{F=1,m_{F}=1}$ hyperfine state of $^{87}${Rb} and $\ket{1,1}$ state of $^{7}${Li}, respectively. The interspecies scattering length can be tuned through the use of Feshbach resonance \cite{PhysRevA.79.012717}. 
A naive estimate of thickness of the vorton $\delta$ is simply thickness of a $^{87}${Rb} vortex, which is given by the healing length $1/\sqrt{8\pi n_1 a_1}$, an order-of-magnitude estimate supported by our numerical calculations.  For definiteness, we pick the parameters which correspond to the first row of \tableref{tab:data} and estimate the healing length to be 0.45 microns, while $\delta = 3.4\mu m$.  The ratio of these two quantities is roughly 7.5, which is seen in \figref{fig:LiRb}.

We can estimate $N_{2}$ by simply multiplying the cross-section of the vorton, its length, and the density of the lithium, $N_{2}\approx (\pi \delta^{2})(2\pi R) n_{2}$. As an example, for Lithium and Rubidium, if the density of lithium is roughly the same as the density of rubidium $(10^{13}/cc)$, and we want $R$ of the order of 10 microns with an $R/\delta$ of 2.5, then $N_{2}$ should be on the order of $10^{4}$.  We show some example situations in \tableref{tab:data}, where we list the results for Li-Rb $(M_{1}/M_{2}=12)$ vorton and K-Rb $(M_{1}/M_{2}=2.12)$ vorton for some realistic values of $n_1$ and $N_2$. These results indicate that the vorton has only a moderate sensitivity 
to the number of particles present. It also indicates that the $R/\delta$ ratio is not large, making the corrections to the approximation we used in computing its energy sizable.  However, as argued, the curvature effects increase $R/\delta$, so the existence of stable vortons is nonetheless supported by this analysis. 

\begin{center}
\begin{table}
    \begin{tabular}{ | l | l | l | l | l | l | l | l | p{5cm} |}
    \hline
    $M_1/M_2$ & $a_{1}/a_{0}$ & $a_{2}/a_{0}$ & $a_{12}/a_{0}$ & $n_1$ [cm$^{-3}$]& $N_2$ & $R/\delta$ & $R$ [$\mu$m] \\ \hline
  12 &  100 & 40 & 5000 & $4\times10^{13}$ & $2\times10^4$ & 2.3 & 7.9 \\ \hline
  12 &  100 & 40 & 5000 & $4\times10^{12}$ & $2\times10^4$ & 2.5 & 19 \\ \hline
  12 &  100 & 40 & 5000 & $4\times10^{13}$ & $2\times10^3$ & 2.5 & 4.4 \\ \hline
  2.12 &  100 & 85.5 & 5000 & $4\times10^{13}$ & $2\times10^4$ & 1.7 & 5.3 \\ \hline
  2.12 &  100 & 85.5 & 5000 & $4\times10^{12}$ & $2\times10^4$ & 1.9 & 12 \\ \hline
  2.12 &  100 & 85.5 & 5000 & $4\times10^{13}$ & $2\times10^3$ & 1.9 & 2.8 \\ \hline
      \end{tabular}
	\caption{$R/\delta$ for different values of parameters for $l=5$. The values of scattering lengths used were obtained from  \cite{Phys.Rev.A70012701}, \cite{Tiecke}, \cite{Science17May2002}. The scattering lengths are measured relative to the Bohr radius $a_{0}$.}
    \label{tab:data}
\end{table}
\end{center}

\section{Production mechanisms}
Up to now we have discussed the properties of static vortons. We now briefly comment on two possible ways to actually create them experimentally.

\underline{Raman Scattering and Gauss-Laguerre Beams}
A possible way to engineer a vorton in cold atoms traps suggested by J.V. Porto\cite{trey} involves three ingredients. 

\begin{figure}[t]
\includegraphics[width=\columnwidth]{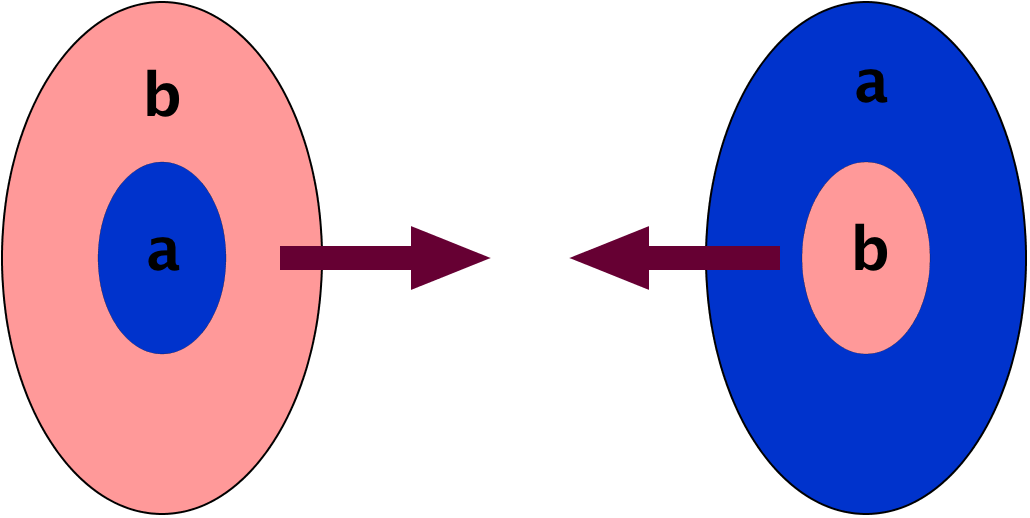}
\caption{Geometry of the two countermoving beams. The arrows denote the direction the beams are moving and the letters ``a'' and ``b'' stand for the two optical frequencies.}
\label{fig:counterbeams}
\end{figure}

First, a closed vortex loop of  $^{87}${Rb} can be created by the use of two counterpropagating light beams, each one with a central circular region with one frequency and an outer annular region with another frequency, as shown in \figref{fig:counterbeams}. In the left-moving beam the inner frequency $\omega_a$ is chosen to match the $D_1$ line of rubidium. The outer frequency $\omega_b$ is chosen to match a transition from the same excited state to another  hyperfine state of $^{87}$Rb. In the counterpropagating beam, the frequencies $\omega_a$ and $\omega_b$ are exchanged. Thus rubidium atoms located in the inner region of the beams will absorb a photon with momentum $\hbar k=\hbar \omega_a/c$ and emit a photon in the opposite direction with nearly the same momentum, due to the stimulation of the second beam. The net effect is that the  $^{87}$Rb atoms in the inner region acquire a momentum $2\hbar k$.  In the annular outer region of the beams the emission and absorption are reversed and the atoms acquire a momentum $-2\hbar k$. 

At the boundary separating the two regions the shearing of the $^{87}$Rb encourages vortex loop formation. The recoil energy of the rubidium atom, for optical photons, is of the order of $10\ \mu K$ so heating may be an issue. This problem might be circumvented  by having only a small fraction of the atoms go through the absorption-emission process. Further scattering of these atoms will distribute its energy and momentum among nearby atoms.

The second ingredient is that the creation of the rubidium loop be accompanied by a change in magnetic field to a value close to the Feshbach resonance between the rubidium and lithium atoms. Such a tuning will entice the system to phase separate, so that the lithium will seek locations with little rubidium: either the boundary of the bulk or the interior of the rubidium vortices. 

Finally, the lithium atoms must acquire a net angular momentum around the vorton before it has time to collapse. This can be accomplished with the use of a Gauss-Laguerre beam. The small values of $l\alt 5$ considered in the numerical examples in \tableref{tab:data} were motivated by the limitations of the Gauss-Laguerre beam technique.  With improved Gauss-Laguerre techniques, a higher $l$ and larger $R/\delta$ could be achieved.

\underline{The Kibble-Zurek Mechanism}
At high temperatures the phase of the bosonic fields $\phi_1$ and $\phi_2$ at different points in space is uncorrelated. At low temperatures, in the superfluid state, the phase has long range correlations. However, following a rapid quench, the phases do not have enough time to correlate and defects form at the regions where different phases meet. This process (the Kibble-Zurek mechanism \cite{0305-4470-9-8-029,Zurek:1985fj}) was demonstrated to create regular vortices in ${}^{87}$Rb\cite{weiler}. In the present case two ingredients should be added. Vorticity needs to be given to the $^{7}$Li using standard methods and a change of the external magnetic field to the Feshbach resonance, in order to impose phase separation.

Any rubidium vortex generated by the standard Kibble-Zurek mechanism will be filled with lithium atoms and any vorticity of the lithium enclosed by the vortex loop will be conserved and will guarantee the stability of the vorton. 

Some simplifying assumptions allows us to estimate the probability of forming a vorton during such a quench. First, we assume that the quench is sufficiently rapid such that the correlation length of the phases of each field is the same as before the quench. Second, we assume that there is little interaction between the two species before the quench (which can be enforced by tuning the interspecies scattering length to zero) but, after the quench, lithium atoms immediately fall into the potential well generated by the core of rubidium vortices. The typical distance between rubidium vortices after the quench is given by the correlation length $\xi_1\sim \hbar/k\sim M_1T/\sqrt{4\pi a_1n_1}$, where $k$ is the typical momentum of a phonon, related to the typical energy of a phonon $\epsilon \sim T \sim c k$, where $c=\sqrt{4\pi a_1 n_1}/M_1$ is the speed of sound. 

Since the orientation of the rubidium vortices is random, a fraction of order one of those vortices will connect to a vortex with the opposite orientation and form a vorton. In those cases, lithium atoms will be trapped in the core of the closed vortex and guarantee its stability as long as there is vorticity in the helium along the closed vortex loop. We can estimate this vorticity by a standard argument\cite{Zurek:1985fj}. The vorticity $l$ is the integral of the phase of $\phi_2$ along the vorton line
\begin{equation}
l = \int d\mathbf{s}.\nabla \theta_2,
\end{equation} where $\phi_2=A_2 e^{i\theta_2}$. The phase of $\phi_2$ changes over a distance of order $\xi_2\sim \hbar/k\sim M_2T/\sqrt{4\pi a_2n_2}$, in exact analogy with $\phi_{1}$. If we assume that the phase of $\phi_2$ is uncorrelated in different regions apart by more than $\xi_2$, we expect the phase of $\phi_{2}$ to change between 0 and $\pi$ a number of times simply given by $2\pi R / \xi_{2}$.  If the phase always changed in the same way, that would be our estimate for $l$.  However, since the phase can change either clockwise or counterclockwise, we expect $l$ to scale like a one-dimensional random walk,
\begin{equation}
l= \int d\mathbf{s}.\nabla \theta_2
\sim
\sqrt{\frac{2\pi R}{\xi_2}},
\end{equation} where $R$ is the vorton radius. The radius of the vorton is of order $R\sim \xi_1$ so we estimate
\begin{equation}
l \sim \sqrt{\frac{2\pi \xi_1}{\xi_2}}
\approx
\sqrt{2\pi \frac{M_1}{M_2}}\left( \frac{a_2 n_2}{a_1 n_1}\right)^{1/4}.
\end{equation}
This estimate suggests that we should pick atoms with very different masses, and that for increased stability created by the Kibble-Zurek mechanism, one should begin with the light species at least as dense as the heavy species, but more dense if possible.  We find that this heuristic estimate suggests, for the case of an equal density  Rb/Li mixture and scattering lengths given by the first line of \tableref{tab:data}, vorticity of \mbox{$l\approx 7$}, a value similar to $l=5$ which was considered in our stability analysis. 

\section{Summary and Outlook}

We considered a cold atom instantiation of vortons, two-species vortex loops.  While vortex loops of a single species will collapse unless supported by the Magnus force, is is possible for vortons to achieve a stable equilibrium radius at rest with respect to the bulk.  Vortons are supported by several astrophysical systems, but have never been observed experimentally in any context.  For realistic laboratory conditions it should be possible to create vortons with radii on the order of micrometers, with $R/\delta$ large enough to provide a believable toroidal geometry.

Aside from the relations to astrophysical systems, such solitons are an interesting aspect of two-species BECs in their own right.  It is an appealing notion to think that such rich structure might be created and manipulated in the laboratory.  The requisite two-species BEC cold-atom traps are a well-established technology, and many two-species pairs have been trapped (see, eg.\ \cite{springerlink:10.1007/s100530050576,PhysRevLett.80.1134,PhysRevA.65.021402}).  While we have focused on a lithium-rubidium system for numerical results, in principle any two-species BEC supports such structures for some choice of vorticity and initial densities.  Aside from computing equilibrium properties, we have discussed two possible mechanisms for crafting vortons in cold atom traps.  Creating such objects would demonstrate their feasibility in other contexts, provide an exciting playground for further investigation, and would widen the possible collaborative avenues between astrophysics and cold atom physics.

\section*{Acknowledgements}
The authors thank Trey Porto for informative discussions. This research was supported by the US Department of Energy under grant number DE-FG02-93-ER40762. 

\bibliographystyle{nar}
\bibliography{vortons_atomic-2011-11-17}

\end{document}